\documentclass[10pt,letterpaper,twocolumn]{article} 

\usepackage{ol2}
\usepackage[draft]{hyperref}
\usepackage{amsmath, amssymb}
\usepackage{overpic}

\newcommand{\tlo}{\tilde{\omega}}

\begin{document}


\twocolumn[ 

\title{Fractional decay of quantum dots in real photonic crystals}

\author{Philip Kristensen,$^{1}$ A. Femius Koenderink,$^2$ Peter Lodahl,$^{1}$ Bjarne Tromborg$^{1}$ and Jesper M\o rk$^{1}$}

\address{
$^1$DTU Fotonik, Technical University of Denmark, \O rsteds Plads building 343, 2800  Lyngby, Denmark
\\
$^2$FOM Institute for Atomic and Molecular Physics, Kruislaan 407, 1098 SJ Amsterdam, The Netherlands \\
$^*$Corresponding author: ptk@com.dtu.dk
}

\begin{abstract}We show that fractional decay may be observable in experiments using quantum dots and photonic crystals with parameters that are currently achievable. We focus on the case of inverse opal photonic crystals and locate the position in the crystal where the effect is most pronounced. Furthermore, we quantify the influence of absorptive loss and show that it is a limiting but not prohibitive effect.
\end{abstract}


] 



\noindent
Spontaneous emission is a resonant process in the sense that an emitter interacts with modes of the electromagnetic field spectrally close to the electronic transition frequency. Moving the emitter to another medium or location at which the field strength of the electromagnetic vacuum modes differ will lead to changes in the spontaneous emission. In most cases the light-matter coupling is weak and the emitter decays exponentially in time with different decay rates at different locations. This is the weak coupling Purcell regime. 

In certain cases the coherent coupling of the emitter to a highly structured electromagnetic vacuum leads to \emph{non exponential} decays. In particular, a regime of so-called fractional decay has been pointed out \cite{Nabiev, John_PRA50_1994}. In this regime, the emitter coherently interacts with modes of low group velocity in such a way that it never fully decays, but rather remains in a superposition of the excited state and the ground state. This may happen in media with rapid variations in the spectral and spatial distribution of electromagnetic modes, as described by the local optical density of states (LDOS) \cite{Sprik_EurophysLett35_1996}. Photonic crystals offer the ability to manipulate the LDOS and change it as compared to the case of a homogeneous medium. The Purcell effect has been shown experimentally for a variety of emitters, e.g, quantum dots (QDs) in inverse opal photonic crystals \cite{Lodahl}, but there is to date no demonstration of fractional decay.


In this Letter we intoduce a practical measure of the degree of fractional decay and use it to investigate decay dynamics of QDs near the band edge of a photonic crystal. For calculations of decay dynamics we follow the approach of Vats \emph{et. al} \cite{Vats_PRA65_2002}. Contrary to the general treatment in \cite{Vats_PRA65_2002} we focus in this Letter on the possible realization of fractional decay using specific and realistic structures. In particular the investigations are based on the actual LDOS of a three dimensional photonic crystal obtained from plane wave calculations and extended to include also effects of absorptive losses, see Fig. 1. Absorption is shown to be a limiting factor and  we present quantitative results showing the degreee of fractional decay achievable for available QDs and practically relevant material loss.
\begin{figure}[htb]
\includegraphics[]{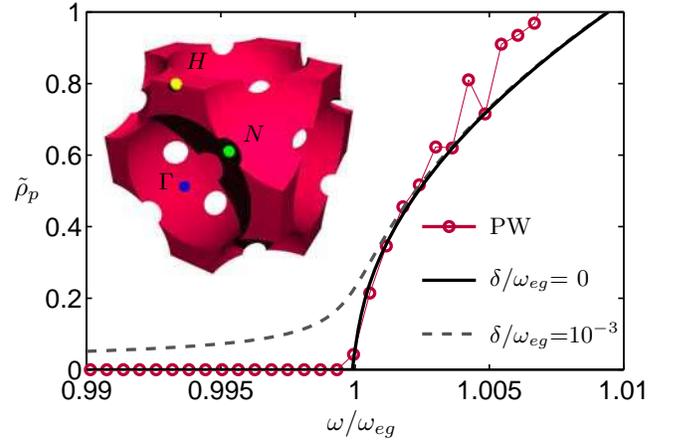}
\caption{\label{Fig:LDOS_2pos_inclAnal}Projected LDOS at the $H$ point of a closed packed Si inverse opal in units of $\omega_{eg}^2/(3\pi^2c^3)$ calculated with plane wave expansion (PW) and analytical approximations in the vicinity of the band edge for two different losses. Inset shows the unit cell of an inverse opal and high symmetry points.}
\end{figure}
%
%

The QD is modelled as an initially excited two-level system with transition frequency $\omega_{eg}$ and we write the general state of the coupled electron-photon system as $|\Psi\rangle=c_e(t)\,|e,0\rangle + \sum_\mu c_{g,\mu}\,(t)|g,\mu\rangle$, where $|e,0\rangle$ denotes the electron in the excited state and no photons and $|g,\mu\rangle$ the electron in the ground state and one photon in mode $\mu$. The time evolution is governed by the Schr\"odinger equation, $i\hbar\,|\dot{\Psi}\rangle  = \hat{H}\,|\Psi\rangle$, with the Hamiltonian $\hat{H}$ in the dipole and rotating wave approximation. Using a Laplace transform \cite{Vats_PRA65_2002} the equations of motion are solved in the frequency domain to yield
\begin{equation}
\tilde{c}_e(\mathbf{r},\tlo)=\frac{1}{\beta G(\mathbf{r},\tlo)-i(\tlo-1)} = \frac{a_{-1}}{\tlo-\tlo_0}+\tilde{c}'(\tlo),
\label{Eq:spectrum}
\end{equation}
where $\mathbf{r}$ is the QD position and $\tlo=\omega/\omega_{eg}$ is the scaled frequency. We have split the spectrum into a pole term with residual $a_{-1}$, pole position $\tlo_0$ and a rest term. The dimensionless parameter $\beta=\Gamma_0/(2\pi\omega_{eg})$ is the vacuum decay rate, $\Gamma_0$, scaled by the transition frequency, and is given as $\beta=q^2\,p^2/(6\,\hbar\,m^2\,\pi^2\epsilon_0\,c^3)$, where $q, p, \hbar, m, \epsilon_0$ and $c$ denote electron charge, momentum matrix element, reduced Planck constant, electron mass, free space permittivity and vacuum speed of light, respectively.  Experimental values range from $\beta\approx10^{-8}$ for InAs QDs \cite{Yu_JPhysChem2005} to $\beta\approx 6\times10^{-8}$ for PbSe QDs \cite{Moreels} with so-called interface defect QDs possibly reaching values of $\beta\approx10^{-6}$ \cite{Andreani}. The function $G(\mathbf{r},\tilde{\omega})$ is given for frequencies above the integration path in the complex plane as
\begin{equation}
G(\mathbf{r},\tlo)\;=\;i\,\tlo\,\int_0^{\tlo_C}\frac{\tilde{\rho}_p(\mathbf{r},x)}{x^2(\tlo-x)}\text{d} x,
\label{Eq:Gintegral}
\end{equation}
in which $\tilde{\rho}_p(\mathbf{r},\tlo)=\rho_p(\mathbf{r},\omega)/\rho_0(\omega_{eg})$ is the ratio of the LDOS to the vacuum LDOS at the emitter frequency. The decay is a resonant process and is governed by the LDOS in a narrow frequency interval around $\tlo=1$. The remaining LDOS, however, does contribute an overall Lamb shift of the spectrum. To model this effect we include an integration using $\rho_p=\rho_0$ for $\tlo<0.95$ and $1.01<\tlo$. A cutoff is chosen at $\tlo_C=10^5$, corresponding to the Compton frequency \cite{Vats_PRA65_2002}. For $0.95<\tlo<1.01$ the integral is carried out using the accurate LDOS $\rho_p=\rho_{BE}$, as calculated below.

The projected LDOS is defined as
\begin{equation}
\rho_p(\mathbf{r},\omega)\;=\;\sum\nolimits_\mu\,|\mathbf{e}_p\cdot\mathbf{E}_\mu(\mathbf{r})|^2\delta(\omega-\omega_\mu),\label{Eq:LDOS}
\end{equation}
where the sum is over all modes of the electromagnetic field indexed by $\mu$ and $\mathbf{e}_p$ is the orientation of the emitter. The functions $\mathbf{E}_\mu(\mathbf{r})=\langle\mathbf{r}\vert\mathbf{E}_\mu\rangle$ denote the spatial and spectral distribution of the modes and are normalized as $\langle\mathbf{E}_{\alpha}\vert\epsilon_R(\mathbf{r})\vert\mathbf{E}_{\beta}\rangle_V=\delta_{\alpha,\beta}$, where $\epsilon_R$ is the relative permittivity and $V$ is the normalization volume. In vacuum the LDOS, Eq. (\ref{Eq:LDOS}), is given as $\rho_0(\omega)=\omega^2/(3\pi^2c^3)$.


Fig. 1 shows a zoom in on the LDOS, $\tilde{\rho}_p(\tilde{\omega})$, at the $H$ point of an Si inverse opal ($\epsilon_R=11.76$) close to the upper edge of the band gap. The LDOS was calculated using a method similar to that of \cite{Busch_PRE58_1998}, corrected for the reduced symmetry of the electric field \cite{Wang_PRB67_2003} and using 169 plane waves and 1232944 $\mathbf{k}$-points distributed over half the full Brillouin zone. The plane wave approach results in a discrete sampling of the LDOS, which effectively limits the slope of the sampled LDOS and leads to incorrect results when used in calculations of fractional decay. For this reason, and in order to include losses in a perturbative way, we analyze analytically the LDOS in the vicinity of the band edge.

The upper band edge in inverse opals is defined by the ninth band at the so-called $X$ point only \cite{Busch_PRE58_1998}. Considering the contribution from just a single band, we follow \cite{Krokhin_PRB53_1996} and rewrite the sum, Eq. (\ref{Eq:LDOS}), as
\begin{equation}
\rho_p(\mathbf{r},\omega)\;=\;\sum_i\frac{V}{(2\pi)^3}\int_{S(\omega)}\,\frac{|\mathbf{e}_p\cdot\mathbf{E}_{\mathbf{k},i}(\mathbf{r}) |^2}{|\nabla\omega(\mathbf{k})|}\,\text{d} \mathbf{k},
\label{Eq:LDOSintegral}
\end{equation}
in which the sum is over two different polarizations and the integration is over the dispersion surface of constant frequency $\omega$ corresponding to the ninth band only. We now expand the integrand in Eq. (\ref{Eq:LDOSintegral}) in powers of $\mathbf{k}$ and carry out the integration. To lowest order the LDOS is given as $\rho_{BE}(\omega)=K_{BE}(\mathbf{r})\sqrt{\omega-\omega_{BE}}$, where $\omega_{BE}$ is the band edge frequency and the band edge parameter, $K_{BE}(\mathbf{r})$, is related to the curvature of the dispersion surfaces and the projected electric field at the point $\mathbf{r}$. The black curve in Fig. 1 illustrates how the square root is indeed the limiting form of the LDOS close to the band edge. Fig. 2 shows values of $K_{BE}$ along lines between symmetry points of the Wigner-Seitz cell of a nearly closed packed Si inverse opal (hole radius per lattice constant, $R/a=0.3436$). The analytical approach allows for the use of only 5 $\mathbf{k}-$points in each direction for the determination of the curvature and 1243 plane waves to achieve convergence\cite{footnote1}.
\begin{figure}[htb]
\includegraphics[]{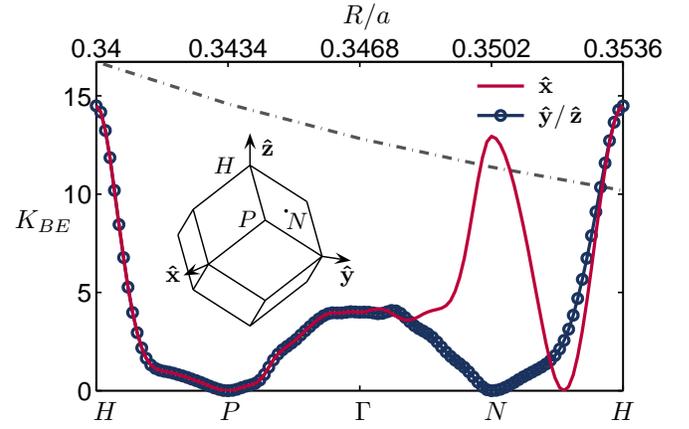}
\caption{\label{Fig:KsVsPosAndRs_Gamma-H-P-Gamma-N-H_2axes}The parameter $K_{BE}$ (in units of $\rho_0(\omega_{eg})/\omega_{eg}^{1/2}$) as a function of position in the Wigner-Seitz cell (shown in the inset) for the three principal emitter orientations. Grey dash-dotted line shows values of $K_{BE}$ at the $H$ point for different $R/a$ (top scale).}
\end{figure}

Introducing loss in the material, $\epsilon_R\rightarrow\epsilon_R+i\,\epsilon_I$, leads to a shift in frequency. For small losses we use first order perturbation theory \cite{Mortensen} to write $\omega=\omega^{(0)}-i\delta$, where $\omega^{(0)}$ is the frequency in the absence of losses and
\[\delta=\,\frac{\omega^{(0)}}{2}\,\frac{\langle\mathbf{E_\mu}|i\,\epsilon_I|\mathbf{E}_\mu\rangle_{C}}{\langle\mathbf{E_\mu}|i\,\epsilon_R|\mathbf{E}_\mu\rangle_V} = \,\frac{\omega^{(0)}\epsilon_I}{2\,\epsilon_R}\,f,
\]
where subsript C denotes the volume of the lossy material only leading to $f=\langle\mathbf{E_\mu}|\epsilon_R|\mathbf{E}_\mu\rangle_{C}/\langle\mathbf{E_\mu}|\epsilon_R|\mathbf{E}_\mu\rangle_V<1$. For nonzero $\delta$ we rewrite the band edge LDOS as \cite{Krokhin_PRB53_1996} 
\[
\rho_{BE}(\mathbf{r},\omega)=K_{BE}(\mathbf{r})\int_{\omega_{BE}}^\infty\sqrt{x-\omega_{BE}}\,\frac{\delta/\pi}{(\omega-x)^2+\delta^2}\,\text{d}x,
\]
which shows that the effect of absorption is to broaden the modes as well as to introduce states below the upper edge of the band gap (dashed curve in Fig. 1).

Using the above expression for $\rho_{BE}(\mathbf{r},\omega)$, the spectrum, $\tilde{c}_e(\mathbf{r},\tlo)$ is calculated from Eqs. (\ref{Eq:spectrum}) and (\ref{Eq:Gintegral}). The temporal evolution is subsequently obtained by transformation back to the time domain. Under this transformation, the pole term in Eq. (\ref{Eq:spectrum}) is conveniently handled analytically, resulting in a decreasing exponential part. The absolute square of the residual denotes the strength of the pole term and is equal to the value of the exponential part at $t=0$. The case of $|a_{-1}|^2=1$ results in $c'(\omega)=0$ and the spectrum consists of only a single pole term. This is characteristic of the weak coupling Purcell regime and the decay is exponential with a decay rate $\Gamma=\Gamma_0\rho_p(\omega_{eg})/\rho_0(\omega_{eg})$. On the other hand, $|a_{-1}|^2<1$ results in a non-zero rest term and consequently a deviation from the Purcell regime and we define $|a_{-1}| ^2<1$ as the condition for fractional decay. The residual depends critically on the light-matter coupling strength, $\beta K_{BE}$, relative to the absorption. The former depends on the QD as well as position in the photonic crystal, cf. Fig. 2.




For a specific example we consider now colloidal PbSe QDs, emitting at $\omega_{PbSe}\approx 1.3\cdot10^{15}\text{s}^{-1}$ ($\beta\approx 5.5\times10^{-8}$ \cite{Moreels}) and placed at the $H$ point in a closed packed Si inverse opal ($K_{BE}\approx 10$). Fig. 3 shows the resulting decay curves for different absorption. For vanishing losses, the population tends to a non-zero value at long times with $|a_{-1}|^2=0.84$. At small finite losses a fractional effect is still visible with $|a_{-1}|^2=0.87$ at an absorption length of $\alpha=3\times10^{-4}\text{cm}^{-1}\;(\delta/\omega_{PbSe}=10^{-10})$ and $|a_{-1}|^2=0.96$ at $\alpha=3\times10^{-5}\text{cm}^{-1}\;(\delta/\omega_{PbSe}=10^{-9})$. We note that absorption in Si at this frequency may be as low as $\alpha\approx 10^{-7}\text{cm}^{-1}\;(\delta/\omega_{PbSe}\approx 10^{-13})$ \cite{Keevers_Solar_1996}. Fig. 3 therefore shows that fractional decay is observable for real QDs in dielectric photonic crystals exhibiting absorptive losses.
\begin{figure}[htb]
\includegraphics[]{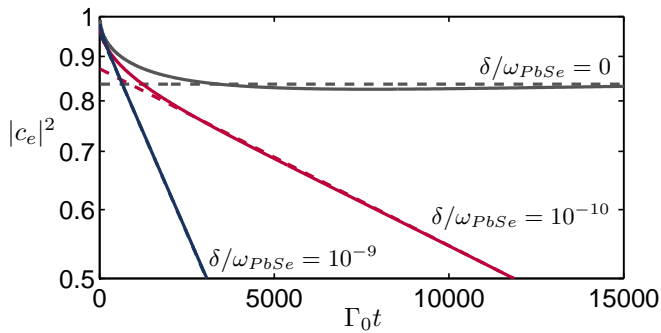}
\caption{\label{Fig:CLEOdecays4}Decay from PbSe QDs at the $H$ point in an Si inverse opal ($\beta K_{BE}=5.5\times10^{-7}$) at detuning $\omega_{BE}/\omega_{PbSe}=1-8.309\times10^{-6}$ and different losses. Dashed curves show exponential parts only.}
\end{figure}
%
%

For a given system the detuning of the emitter relative to the band edge defines the exact modes, and hence group velocity, of the emitted light. Therefore, the residual depends also on the detuning and we define the parameter $D_f$ as the minimum value of $|a_{-1}|^2$ for optimized detuning. In Fig. 4 we show $D_f$ as a function of $\beta K_{BE}$. The curves were obtained for each $\beta K_{BE}$ by varying the detuning until a minimum was found. The figure shows that a profound degree of fractional decay is possible for a range of experimentally relevant material parameters.
\begin{figure}[htb]
\includegraphics[]{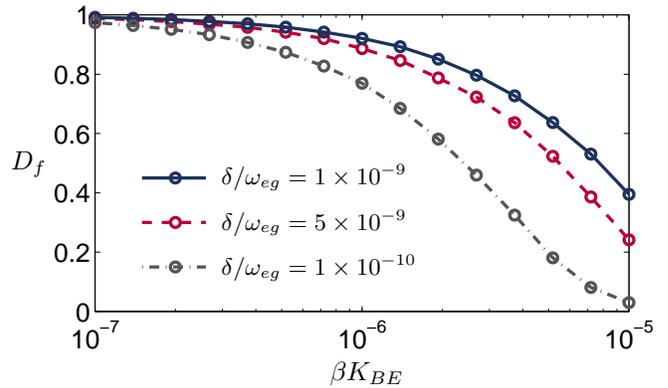}
\caption{\label{Fig:minabsress_beta1em7_varloss}Degree of fractional decay obtainable versus light-matter coupling strength, $\beta K_{BE}$, at different absorption.}
\end{figure}

In conclusion we have used an analytic expression to the band edge LDOS to investigate fractional decay dynamics in inverse opals. The analysis has revealed the position in the crystal that is most suitable for observation of fractional decay. Furthermore, we have extended the analysis to include absorptive losses and calculated the degree of fractional decay obtainable for given losses and light-matter coupling strengths. The analysis shows that absoption has a limiting but not prohibitive effect and that fractional decay may be possible to achieve using, e.g., PbSe QDs in Si inverse opals.

Enlightening discussions with Willem Vos and Niels Asger Mortensen are greatly appreciated.



\begin{thebibliography}{99}
\bibitem{footnote1}
Values along the $\Gamma-H$ route showed poor convergence and have been left out since they are irrelevant to the conclusions in this Letter.
\end{thebibliography}
\end{document}